\documentclass[prl,twocolumn,showpacs]{revtex4}
\usepackage[dvips]{graphicx}
\usepackage{color}
\usepackage{amsmath}
\usepackage{amsfonts}
\usepackage[dvips]{graphicx}

\begin{document}
\newcommand{\be}{\begin{equation}}
\newcommand{\ee}{\end{equation}}
\newcommand{\bea}{\begin{eqnarray}}
\newcommand{\eea}{\end{eqnarray}}
\title{Complete, Single-Horizon Quantum Corrected Black Hole Spacetime}

\author{Ari Peltola}
\email[Electronic address: ]{a.peltola-ra@uwinnipeg.ca} 
\author{Gabor Kunstatter}
\email[Electronic address: ]{g.kunstatter@uwinnipeg.ca} 
\affiliation{Department of Physics and Winnipeg Institute for Theoretical Physics, The University of Winnipeg, 515 Portage Avenue, Winnipeg, Manitoba. Canada. R3B 2E9}

\begin{abstract} 
We show that a semi-classical polymerization of the interior of Schwarzschild black holes gives rise to a tantalizing candidate for a non-singular, single horizon black hole spacetime. The exterior has non-zero quantum stress energy but closely approximates the classical spacetime for macroscopic black holes. The interior exhibits a bounce at a microscopic scale and then expands indefinitely to a Kantowski-Sachs spacetime. Polymerization therefore removes the singularity and produces a scenario reminiscent of past proposals for universe creation via quantum effects inside a black hole.
\end{abstract}


\maketitle

\section{Introduction}
Recent work suggests that Loop Quantum Gravity (LQG) may be capable of resolving the singularities that are inevitable in classical general relativity. Because of the inherent difficulty in solving the complete system, the focus has been 
on dimensionally reduced, mini-superspace models of quantum cosmology \cite{lqc} and  spherically symmetric black hole spacetimes \cite{Ashtekar05, modesto06, boehmer07, pullin08}. As observed in \cite{Ashtekar05} and \cite{modesto06}, for example, singularity avoidance in these two cases is deeply connected because the interior Schwarzschild spacetime coincides with the homogeneous anistropic Kantowski-Sachs cosmology.  While significant progress has been made, there is as yet no clear evidence as to which theory of quantum gravity will ultimately be proven correct, LQG, string theory, or perhaps something else.  Nor is there, to the best of our knowledge, a rigorous and unambiguous path from the full loop quantum gravity theory to the quantization techniques used in the mini-superspace models. One particularly fruitful technique that has been used recently to great effect \cite{boehmer07, pullin08} is a semi-classical polymerization that preserves aspects of the underlying discreteness of spacetime suggested by LQG but considers the limit in which quantum effects are vanishingly small. Different polymerizations can give qualitatively different regularized spacetimes, so that it is of great interest to examine more fully  a wider class of models and methods. 

In the following, we describe quantum corrections that arise from the semi-classical polymerization of the interior of generic black holes in a family of theories known collectively as generic 2-D dilaton gravity. Of prime importance for the present work is that this family includes spherically symmetric black holes in spacetime dimension three or higher. We investigate two different polymerization schemes, and show that the results differ qualitatively: in one case the resulting non-singular spacetime generically has only a single horizon while in the second there are multiple horizons. 
In our considerations, the polymerization scale is taken to be a constant but we note that in the context of loop quantum cosmology, as well as in some black hole scenarios \cite{nelson}, a dynamical polymerization scale has been considered.

Our key result is the analytic solution of the semi-classical equations that is obtained in 4-D when only area is polymerized. This solution can be extended analytically to a complete non-singular spacetime with only a single horizon. This has the advantage over other candidates for loop quantum corrected black holes \cite{boehmer07, pullin08} of avoiding the problem of mass inflation \cite{mass_inflation} normally associated with inner horizons.  The exterior has non-zero quantum stress energy but closely approximates the classical spacetime for macroscopic black holes. There are two interior regions, one in the past and one in the future. Both exhibit a bounce at a microscopic scale and then asymptote (one in the infinite past and the other in the infinite future) to a non-singular product Kantowski-Sachs \cite{kantowski-sachs} type cosmological spacetime containing an anisotropic fluid, with product topology of a spacelike R and an expanding 2+1 positive curvature FRW cosmology. The polymer dynamics thus drives the system into an asymptotic interior end-state that is not a small correction to the classical spacetime. In the limit that the polymerization scale goes to zero, the interior cosmological regions ``pinch off'' leaving behind the standard singular Schwarzschild interior. The complete, non-singular semi-classical spacetime is suggestive of past proposals for ``universe creation'' inside black holes \cite{frolov90}.

\section{Classical Theory}
Our formalism  begins with the most general (up to point reparametrizations) $1+1$-dimensional, second order, diffeomorphism 
invariant action that can be built from a 2-metric $g_{\mu \nu}$ and a scalar $\phi$ \cite{gru,dil1}:
\be \label{eq:act} S[g,\phi ] = \frac{1}{2G}\int d^2x \sqrt{-g}\, \bigg( \phi R(g)
	+ \frac{V(\phi)}{l^2}\bigg),
\ee
where $l$ is a positive constant with 
a dimension of length and $G$ is the dimensionless two-dimensional Newton's constant. This action provides a convenient representation of spherically symmetric black hole spacetimes in $D\equiv n+2$ dimensions. It obeys a generalized Birkhoff theorem \cite{dil1} with general solution:
\be \label{eq:ds} ds^2 = -\big[ 2lGM-j(\phi )\big]^{-1}l^2 d\phi^2+\big[ 2lGM-j(\phi )\big]dx^2.
\ee
where $j(\phi )$ satisfies $dj/d\phi=V(\phi )$. For our purposes, it is convenient to assume that $j(\phi)\rightarrow 0$ when \mbox{$\phi\rightarrow 0$}. The integration constant $M$ is the Arnowitt-Deser-Misner (ADM) mass, and we take $M>0$.
For monotonic functions $j(\phi )$ the solution contains precisely one Killing horizon \cite{dil1} at $\phi_\text{H}$, such that $j(\phi_\text{H}) = 2lGM$. The metric (\ref{eq:ds}) and dilaton can be related to a spherically symmetric metric in $D$ dimensions as follows:
\be 
ds^2_\text{phys}= \frac{ds^2}{j(\phi )} + r^2(\phi) d\Omega^2_{(D-2)},
\label{eq:physical_metric}
\ee
where $d\Omega^2_{(D-2)}$ is the line element of the unit $(D-2)$-sphere. Eq. (\ref{eq:physical_metric}) corresponds in particular to the interior of the 4-D Schwarzschild solution with the identifications $2Gl^2=G^{\text{(4)}}$, $V=1/(2\sqrt{\phi})$ and $\phi=r^2/(4l^2)$.

In order to address the question of singularity resolution in the semi-classical polymerized theory, we restrict to homogeneous slices with metric parametrization
\be \label{eq:adm}
ds^2 = e^{2\rho(t)}\big( -\sigma^2(t)dt^2+dx^2\big).
\ee
After suppressing an irrelevant integration over the spatial coordinate, the resulting action is that of a parametrized system:
\be I=\frac{1}{2G}\int dt\Big(\Pi_\rho\dot{\rho}+\Pi_\phi\dot{\phi} +\sigma \mathcal{G}\Big),
\label{eq:action}
\ee
where a dot denotes a derivative with respect to the time coordinate $t$ and the (Hamiltonian) constraint is
\be 
\mathcal{G}=G\Pi_\rho \Pi_\phi +e^{2\rho} \frac{V}{2l^2G}\approx 0.
\label{eq:constraint}
\ee
In spherically symmetric 4-D spacetime, this Hamiltonian can be converted to the LQG Hamiltonian of \cite{pullin08} by a simple point canonical transformation.

The simplicity of the Hamiltonian in (\ref{eq:constraint}) makes it rather straightforward to find analytic solutions for the components of the physical metric. These solutions depend on two parameters, the ADM mass $M$, and its canonical conjugate $P_M$. In the full (inhomogeneous) spherically symmetric theory the latter corresponds to the Schwarzschild time separation of spatial slices \cite{kuchar94}. In the present case, the arbitrariness of $P_M$ represents the residual invariance of the theory under rescalings of the Schwarzschild ``time'' coordinate $x$.

\section{Semi-Classical Polymer Approach}
In the polymer representation of quantum mechanics \cite{ash,halvorson} one effectively studies
the Hamiltonian dynamics on a discrete spatial lattice. 
In many cases the full polymer theory is rather challenging to analyze
but fortunately one can get interesting results  by investigating the 
semi-classical limit of the theory, which corresponds formally to the limit in which 
quantum effects are small ($\hbar\to 0$), but the polymerization scale $\mu$ stays finite. 
This approximation is the basis for recent analyses of black hole interiors \cite{boehmer07,pullin08}. It can be derived  by studying the action of the fully quantized operators on coherent states and expanding in the width of the states \cite{husain:semiclass,ding08}. The end result is to simply replace the classical momentum variable $p$ in the classical Hamiltonian function by $\sin (\mu p)/\mu$. After the replacement, one studies the \hbox{(semi-)classical} dynamics of the resulting polymer Hamiltonian by means of standard techniques. We assume in the following that this effective description is valid without further state-dependent corrections, at least for states that approach semi-classical behaviour at scales large compared to $\mu$.

We first polymerize only the generalized area variable $\phi$. While this may seem {\it ad hoc}, it is perhaps not unreasonable to introduce a fundamental discreteness for the geometrical variable that corresponds to area in the spherically symmetric theory while leaving the coordinate dependent conformal mode of the metric continuous.  This approach is motivated by (but not derived from) full LQG, where it is the area operator that is naturally discretised. For completeness we will subsequently illustrate the result of polymerizing both variables. Details of both polymerizations will be presented elsewhere \cite{AK:prep}. Note that we choose to work with a state-independent (constant) polymerization scale, despite the fact that in the context of loop quantum cosmology consistency with predictions requires a state-dependent discreteness scale\cite{lqc}. Here we choose the simplest approach that produces reasonable semi-classical behaviour and leave the study of state-dependent polymerization scales for future research.

The partially polymerized Hamiltonian constraint is:
\be \label{eq:polyH}\mathcal{G} =  G \Pi_\rho\frac{\sin (\mu\Pi_\phi)}{\mu}
  + e^{2\rho}\frac{V(\phi)}{2l^2G}\approx 0.
\ee
In this equation $\mu$ has a dimension of length, which means that $\phi$ has a discrete polymer structure with edge length of $\mu/l$. The essence of the  singularity resolution mechanism is evident from the equation of motion for $\phi$:
\begin{eqnarray}
  \frac{\dot{\phi}}{G\sigma}&=& -\Pi_\rho \cos(\mu\Pi_\phi).
\label{eq:phidot}
\end{eqnarray}
$\dot{\phi}$ now vanishes at two turning points: the ``classical'' turning point  $\Pi_\rho=0$ and semi-classical turning point: $\cos(\mu\Pi_\phi)=0$. The former condition turns out to be satisfied at the horizon as expected while the latter occurs first at a microscopic scale proportional to $\mu$.

The solution to the corresponding H-J equation is: 
\be \label{eq:HJF} S=-\frac{\alpha}{4G}\, e^{2\rho}+\frac{l}{\mu}\int \arcsin 
	\bigg( \frac{\mu V}{\alpha lG}\bigg)\, d\phi +C,
\ee
where $\alpha$ and $C$ are constants.  For convenience, we shall take $\alpha >0$. 
The expression for $\Pi_\phi$ now reads:
\be
\label{eq:pi1}\Pi_\phi =\frac{1}{l}\frac{\partial S}{\partial \phi } = 
	\frac{1}{\mu}\arcsin \bigg( \frac{\mu V}{\alpha lG}\bigg). 
\ee
We concentrate henceforth on spherically symmetric gravity for which $V\propto\phi^{-1/n}$. In this case $\phi$ has an minimum given by $\mu V(\phi_{\text{min}})=\alpha lG$. Note that $\phi_\text{min}$ is located at the roots of the cosine function as expected from (\ref{eq:phidot}). 

To find the solutions, we  require:
\be \label{eq:beta2} \frac{\partial S}{\partial\alpha}=-\beta,
\ee
where $\beta$ is again a constant of motion that is conjugate to $\alpha$. The constants $\beta$ and $\alpha$ are related to the usual canonical pair $(M,P_M)$ by a simple canonical transformation. Eq. (\ref{eq:beta2}) yields a solution for $e^{2\rho}$ in terms of $\phi$ and the two constants $\alpha$ and $\beta$ that are determined by initial conditions. The solution is: 
\be \label{eq:rho soln} \frac{1}{4G}\, e^{2\rho}+\frac{l}{\mu} I^{(n)}(\phi ) = \beta ,
\ee
where 
\bea
I^{(n)}&:=& \frac{\epsilon}{\alpha}\int\frac{V}{\sqrt{\left(\alpha lG/\mu\right)^2-V^2}}\,d\phi \, .
\label{eq:I}
\eea 
In (\ref{eq:I}), the value of $\epsilon=\pm 1$ depends on a branch of $\mu\Pi_\phi$. The upper sign is valid in the branches where the cosine function is positive, which include the principal branch $(-\pi/2,\pi/2)$, whereas the lower sign is used elsewhere.

One now has a complete solution in terms of a single arbitrary function that must be fixed by specifying a time variable. It is illustrative to write the physical metric using $\phi$ as a coordinate, which corresponds to the area of the throat in the interior of the extended Schwarzschild solution of the unpolymerized theory. Suppressing the angular part, one obtains 
\be ds^2_\text{phys}=\frac{1}{j(\phi)}\Big(\frac{-4l^2d\phi^2}{\alpha^2
e^{2\rho}[1-(\mu V/(\alpha lG))^2]}+e^{2\rho}dx^2\Big), 
\ee
which shows that the solution has a horizon at $\phi=\phi_{\text{H}}$ for which $e^{2\rho}=0$.

\begin{figure}[htb!]
\begin{center}
\includegraphics{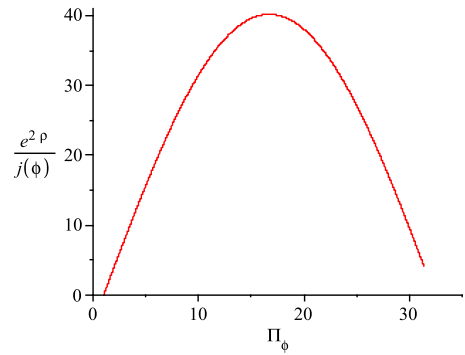}
\caption{The physical conformal mode in 4-D spacetime. The calculations use $\alpha=G^\text{(4)}=M=1$ and $\mu=0.1$.} 
\label{fig:bounce}
\end{center}
\end{figure}

Taking $\Pi_\phi$ as the time variable, one can deduce, generically, the following qualitative time evolution. One starts at the horizon $\phi=\phi_\text{H}$ , with corresponding initial $\Pi_\phi^\text{H}$ given by (\ref{eq:pi1}). Without loss of generality, we can take $\mu\Pi_\phi^\text{H}$ to be in the principle branch of the arcsin function, and because $\phi$ is positive, it takes values between $(0,\pi/2)$. As $\mu\Pi_\phi$ increases, $\phi$ decreases until it reaches its minimum value at $\mu\Pi_\phi =\pi/2$. After that $\phi$ increases as $\Pi_\phi$ increases. When $\mu\Pi_\phi$ is in the range $(\pi/2, \pi)$, $\epsilon$ in (\ref{eq:I}) necessarily changes sign, and one can verify that after the bounce $e^{2\rho}$ does not vanish again, but the throat area expands to $\phi\to\infty$ in finite coordinate time $\Pi_\phi$. The expansion takes an infinite amount of proper time so that our semi-classical polymerization has produced a solution that avoids the singularity but does not oscillate. The time evolution of the physical conformal mode, $e^{2\rho}/j(\phi)$, is illustrated in Fig. \ref{fig:bounce}.

We now contrast the above behavior with that of the fully polymerized theory. The Hamiltonian constraint is:
\be \label{eq:fullyH}\mathcal{G} =  G \frac{\sin(\mu\Pi_\rho)}{\mu}\frac{\sin (\mu\Pi_\phi)}{\mu}
  + e^{2\rho}\frac{V}{2l^2G}\approx 0.
\ee
Going through precisely the same Hamilton-Jacobi analysis as before, we are lead to the following expression for the
conformal mode of the metric:
\be
e^{2\rho}= \frac{2lG}{\alpha\mu}\sin\big( 2\mu\alpha\beta/l-2\alpha I^{(n)}(\phi)\big),
\label{eq:fullrho}
\ee
where $I^{(n)}$ is given in Eq. (\ref{eq:I}). All other $\phi$-dependence is unchanged so that the net change from the partially polymerized case is that the conformal mode is now an oscillating function of $\phi$. There will now be inner horizons whenever the sin function vanishes, giving rise to a qualitatively different quantum corrected spacetime. In fact the number of horizons  will vary depending on the relative magnitude of $M$ and the quantity $\mu /\alpha$ \cite{AK:prep}.

\section{Four-Dimensional Schwarzschild Black Hole}
We now write the 4-D partially polymerized metric in terms of the radius $r$ and the Schwarzschild ``time'' $x$:
\bea
\!\!\!\!\!\!\!\!\!\!\!ds^2_\text{phys}\!\!&=&\! -\frac{dr^2}{\Big(\frac{2MG^{(4)}}{r} - \epsilon\sqrt{1-\frac{k^2}{r^2}}\Big)\Big(1-\frac{k^2}{r^2}\Big)} \nonumber\\
& &+  \bigg(\!\frac{2MG^{(4)}}{r} - \epsilon\sqrt{1-\frac{k^2}{r^2}}\,\bigg)\!\bigg(\!\frac{2dx}{\alpha}\!\bigg)^{\!\!\!2} \!+\!r^2d\Omega^2.
\label{eq:four_metric}
\eea
In the above, $M \equiv\alpha^2 \beta/(2l)$ and $k\equiv \mu/(\alpha G)$.  As per our earlier claim, $P_M=2l/\alpha$ completes the canonical transformation from the pair $(\alpha,\beta)$  to $(M,P_M)$, and hence the conjugate $P_M$ does indeed rescale the $x$ coordinate. These rescalings do affect the bounce radius $k$ which is a consequence of the fact that the introduction of the discrete scale has broken the scale invariance of the theory.

The metric (\ref{eq:four_metric}) has remarkable properties. There is a single bifurcative horizon at  $r_\text{H} := \sqrt{(2MG^{(4)})^2+k^2}$ which exhibits a quantum correction due to the polymerization. The solution evolves from the horizon at $r_\text{H}$ to the minimum radius $k$ in finite proper time, and then expands to $r=\infty$ in infinite proper time. As the throat expands in the interior, the metric approaches: 
\be
ds^2_\text{phys} = -\frac{dr^2}{1+2MG^{(4)}/r}+\Big(1+\frac{2MG^{(4)}}{r}\Big)dx^2,
\label{eq:asympt_interior}
\ee
where we have absorbed $2/\alpha$ into $x$ and suppressed the angles. This asymptotic interior solution does not obey the vacuum Einstein equations, but has non-vanishing stress tensor with $T^{\,r}_r = T^{\,x}_x \propto -1/r^2$ which does not depend on $k$. This corresponds to an anisotropic perfect fluid that has been recently considered in a model of the Schwarzschild interior \cite{culetu}. 

It is possible to continue the metric (\ref{eq:four_metric}) analytically across the horizon to the exterior region. The validity of this extension is an open question given that our chosen foliation does not extend to the exterior, but the procedure seems natural in the present context and has been used before to construct a complete semi-classical black hole spacetime.  In the present case, the resulting black hole exterior has interesting and physically reasonable properties: in particular, it is asymptotically flat and closely approximates a Schwarzschild black hole of mass $M$.

\begin{figure}[htb!]
\begin{center}
\includegraphics{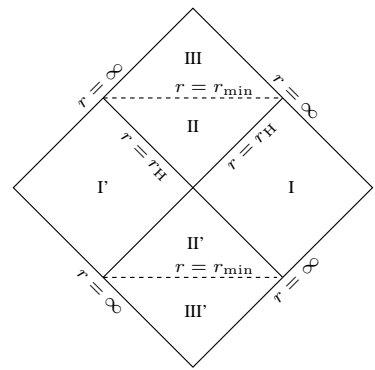}
\caption{Conformal diagram of the partially polymerized Schwarzschild spacetime. }
\label{fig:conformal}
\end{center}
\end{figure}

The fact that $r=k$ in (\ref{eq:four_metric}) is a coordinate singularity can be explicitly illustrated by the coordinate transformation $r=k\cosh(y)$ \cite{ftnote}. The resulting metric is regular at the bounce $y=0$ and has a horizon at $\sinh(y_\text{H})=2MG^{(4)}/k$. For large $y$, the metric describes the exterior of the black hole, while the asymptotic interior region corresponds to $y\to -\infty$. The conformal diagram of the complete spacetime is shown in Fig. \ref{fig:conformal}. It includes two exterior regions (I and I'), the black hole and the white hole interior regions (II and II'), and two ``quantum corrected'' interior regions (III and III'). The classical singularity is replaced by a bounce at $r=r_\text{min}$ and subsequent expansion to $r=\infty$.

The Ricci and Kretschmann scalars are everywhere non-singular and vanish rapidly in the exterior (large, positive $y$). 
A calculation of the Einstein tensor reveals that while the solutions violate the classical energy conditions, the violations are of order $k^2/r^4$ and hence vanish far from the bounce radius $r=k$. This means in particular that the exterior is endowed with non-zero quantum stress energy that is vanishingly small for macroscopic black holes ($r_\text{H}\gg k$) so that the Schwarzschild solution is well approximated everywhere outside the horizon. This also leads naturally to a quantum corrected Hawking temperature of order $\mathcal{O}(k^2/M^3)$ as well as a logarithmic correction to the Bekenstein-Hawking entropy \cite{AK:prep}.

The interior spacetime on the other hand describes an expanding anisotropic cosmology with stress energy that approximately satisfies the energy conditions but does not vanish far from $k$, nor does it vanish in the limit that $k \to 0$. Instead, the asymptotic region ``pinches off'' in this limit at the curvature singularity at $r=0$, leaving behind the standard, complete but singular Schwarzschild spacetime and two disconnected, time-reversed copies of the (singular) cosmological spacetime.

\section{Conclusion}
We have presented an intriguing candidate for a complete, non-singular quantum corrected black hole spacetime. This spacetime was derived by the semi-classical polymerization of only the area in the interior of spherically symmetric black hole spacetimes. Our treatment neglects polymeric corrections to the conformal mode of the metric, but this somewhat speculative procedure is perhaps justified by the interesting results that emerge. The singularity is resolved at a bounce radius determined by the polymerization scale and the exterior black hole spacetime has small, but non-zero quantum-energy. Remarkably, the solution in the interior does not oscillate, but instead re-expands indefinitely to a Kantowski-Sachs spacetime with anisotropic fluid stress-energy that is non-vanishing in the limit that the polymerization scale vanishes. The generation via polymerization of an interior cosmology is reminiscent of earlier work that explored universe creation in black hole interiors \cite{frolov90}.

One may also note that the solution does not reduce to Minkowski space in the limit that the Schwarzschild mass $M$ goes to zero. In fact, Eq. (\ref{eq:four_metric}) shows that there is still a horizon in this limit, located at the bounce radius $r=r_\text{min}$. While it is tempting to speculate about quantum remnants, it must be remembered that the semi-classical approximation employed here will likely break down for microscopic black holes. This is certainly worthy of further investigation.

{\bf Note Added:} After this paper was completed, a paper by Modesto \cite{modesto08} appeared which presents an interesting and thorough analysis of the complete LQG corrected, multiple horizon 4-D black hole spacetimes that emerge from a generalization of the procedure in \cite{pullin08}.

\section*{Acknowledgements}

We thank J. Ziprick, R. Daghigh, J. Gegenberg, J. Louko, the theory group at CECS and H. Maeda in particular for useful discussions.  GK acknowledges the kind hospitality of the University of Nottingham, the University of New Brunswick and CECS where parts of this work were carried out. This work was supported in part by the Natural Sciences and Engineering Research Council of Canada.

\end{document}